\documentclass[12pt,preprint]{aastex}
\newcommand{\myemail}{lychiang@asiaa.sinica.edu.tw}

\usepackage{float}
\usepackage{amsmath}
\usepackage{graphicx}
\usepackage{epsfig}
\usepackage{subfigure}
\makeatletter

\def\alm{a_{\l m}}

\def\glesp{G{\sc lesp }}
\def\healpix{H{\sc ealpix }}

\def\wmap{{\rm WMAP }}

\def\etal{{et al.}}
\def\l{{\ell}}

\def\ylm{Y_{\l m}}

\def\sumlm{\sum_{\l m}}

\slugcomment{Submitted to Astrophysical Journal}

\shorttitle{Asymmetry and non-random orientation of the inflight beam in the \wmap data
}
\shortauthors{Lung-Yih Chiang}

\begin{document}
\title{Asymmetry and non-random orientation of the inflight effective beam pattern in the \wmap Data}
\author{Lung-Yih Chiang}
\affil{\asiaa}

\email{\myemail}

\newcommand{\asiaa}{{Institute of Astronomy and Astrophysics, Academia Sinica, P.O. Box 23-141, Taipei 10617, Taiwan}}

\begin{abstract}
The anomaly against the Gaussianity in the WMAP data was alleged to be due to “insufficient handling of beam asymmetries”. In this paper we investigate this issue and develop a method to estimate the shape of the inflight effective beam, particularly the asymmetry and azimuthal orientation. We divide the whole map into square patches and exploit the information in the Fourier space. For patches containing bright extra-galactic point sources, we can directly estimate their shapes, from which the inflight effective beam manifests itself. For those without, we estimate the pattern via perturbing the phases and directly from the Fourier amplitudes. We show that the inflight effective beam convolving the signal is indeed non-symmetric for most part of the sky, and it's not randomly oriented. Around the ecliptic poles, however, the asymmetry is smaller due to the averaging effect from different orientations of the beam from the scan strategy. The effective beam with significant asymmetry is combing with almost parallel fashion along the lines of Ecliptic longitude. In the foreground-cleaned ILC map, however, the systematics caused by beam effect is significantly lessened. 

\end{abstract}

\section{Introduction}
The measurement of the cosmic microwave background (CMB) by NASA Wilkinson Microwave Anisotropy Probe (\wmap) \citep{wmap1yrresult,wmap1yrcos,wmap3yrtem,wmap3yrcos,wmap5yrresult,wmap5yrcos,wmap7yrresult,wmap7yrcos,wmap9yrresult,wmap9yrcos} and ESA Planck Surveyor \citep{planckmission,planckercsc,planckcluster,planckresult,planckcos} has enabled us to probe cosmology in high precision. The CMB signal is observed through the convolution of an antenna beam, an effect, and possible systematic error, which must be carefully
treated in the data analysis. The main beam of the \wmap are shown not azimuthally symmetric about the line of sight, but rather elliptical (see Table 1) due to the fact that they cannot be all put on the center of the focal plane, but it was assumed that the inflight effective beam convolving the signal should be averaged to become symmetric and circular after each pixel on the map having more than at least 400 hits during the 1-year observation \citep{wmap1yrbeam}. Prior to \wmap data release, the issue of beam asymmetry in full-sky CMB experiment is already discussed in \citet{burigana,fosalba}. The issue of asymmetry of the \wmap beam is later tackled in \citet{wmap9yrresult} with a new map-making procedure, which resultantly deconvolves the beam sidelobes to produce maps with the true sky signal convolved by symmetrized beams. Whilst the issue of beam asymmetry is considered on the estimation of the CMB angular power spectrum \citep{mitra1,mitra2} and is demonstrated to have less effect \citep{wmap3yrtem,norwayasym}, it poses a serious issue on statistical isotropy, the foremost example being the quadruople and octupole alignment \citep{wmap7yranom} or anomaly in the \wmap result such as low quadrupolar power \citep{cambridgeasym, wmap9yrresult}. It is because systematic alignment in orientation of even a mildly-asymmetric beam can result in large-scale anisotropy. 

To investigate this issue, we would like to see what {\it effective} beam pattern is convolving the signal. Note that for each pixel the asymmetric inflight beam with different azimuthal orientation (around the line-of-sight direction) convolves the underlying signal thousands of times with some pointing uncertainty, so we are particularly interested in the asymmetry and azimuthal orientation of the {\it effective} elliptical main beam which is from the collective effect of the inflight beam. Theoretically, one can gather all the inflight beam information through the time-ordered data and formulate the effective beam pattern for each point on the sky, but nevertheless it is difficult to get the effective beam pattern in the final product such as the Internal Linear Combination (ILC) map due to the complex noise properties. Throughout this paper we define $r\equiv r_{\rm maj}/r_{\rm min}$, where $r_{\rm maj}$ and $r_{\rm min}$ are the major and minor axis of an elliptical shape, respectively,  and  the asymmetry of the beam as $R\equiv r -1$.

In this paper we develop a method that can reveal the asymmetry and azimuthal orientation of the effective main beam in different small patches of the sky, based on the flat sky approximation. There are two ways the effective beam convolving the signal manifests itself in the signal. One is through the shape of the bright extragalactic  point sources, the other is in the Fourier amplitude, the latter of which is particularly useful for the processed map such as the ILC map.

This paper is arranged as follows. In Sec.2 we introduce the Fourier method and in Sec.3 we test the accuracy and consistency of the method. We then employ the methods on \wmap data in Sec.4, and the Conclusion and Discussion is in Sec.5.

\section{Manifestation of the orientation and asymmetry of the inflight effective beam in the Fourier space} 
In CMB experiment, the temperature measured in the sky can be written as $S= (C+F)\star B +N$, where $C$ and $F$ are the CMB and foreground signal, respectively, $B$ is the effective beam convolving the signal, $N$ is the noise, and the star sign $\star$ denotes convolution. The convolving beam was often assumed azimuthally symmetric and thus $B (\theta)=\sum_\l (2\l+1/4\pi) b_\l P_\l(\cos \theta)$ (i.e no $m$ dependence). and in terms of spherical harmonic coefficients,  
\begin{equation}
S(\theta,\phi)=\sumlm{\alm b_{\l}}\ylm(\theta,\phi) +N(\theta,\phi),
\label{sphar}
\end{equation}    
where the $\ylm$ is the spherical harmonics and $\alm$ is the spherical harmonic coefficients of the signal from the sky. Due to the scan strategy, the inflight effective beam has different asymmetry and orientation to different line of sight. If we are to examine the anisotropies caused by the inflight beam asymmetry, one has to resort to its effect in small patches, where the beam geometry can be described in the flat-sky approximation. Within this approximation, the signal measured is now
\begin{equation}
S({\bf k})= \int d^2 {\bf r}[ B({\bf r})\star T({\bf r}) +N({\bf r}) ]\exp({2\pi i {\bf k}\cdot {\bf r}})= \sqrt{C({\bf k})} \exp[i \Psi({\bf k})] 
\label{gen}
\end{equation}
To estimate the inflight effective beam, we assume the main beam is of an elliptical Gaussian shape, which can be expressed in Cartesian coordinate
\begin{equation}
B(x,y)=\frac{1}{2 \pi \sigma_x \sigma_y} \exp\left[-\frac{(x \cos\gamma +y \sin \gamma)^2}{2\sigma_x^2}-\frac{(-x \sin \gamma +y \cos \gamma)^2}{2\sigma_y^2}\right], 
\end{equation}
where $\gamma$ is the orientation against the $x$-axis. 

We have tested the \wmap beam asymmetry and listed in Table 1. The \wmap beam maps are from observation of Jupiter and we list the A and B side for K, Ka, Q1, V1 and W1 Differencing Assembly. The high asymmetry is due to the feeds being away from the primary focus. It is significantly lessened from multiple observations with different orientation in each pixel.

\begin{table}[t]
\begin{tabular}{|c||c|c|c|c|c|}
\hline
R & K & Ka & Q1 & V1 & W1  \\
\hline
A   & 0.4258   & 0.2950 &  0.3217   & 0.1409 &  0.0970\\
\hline
B   & 0.4192   & 0.2913 &  0.3305   & 0.1366 & 0.0918\\
\hline
\end{tabular}
\caption{\wmap beam asymmetry $R$. The \wmap beam maps are from Jupiter observation and we list the A and B side of the focal planes for K, Ka, Q1, V1 and W1 band. The asymmetry is high because the feeds are away from the primary focus.}
\end{table}

To estimate the effective beam, we introduce two methods, both utilizing the information in Fourier domain. In this paper we assume that the effective beam is slowly rotating across the sky, thus for a small patch of sky the effective beam convolving the signal shall have a fixed orientation. 

\begin{figure}
\centering
\epsfig{file=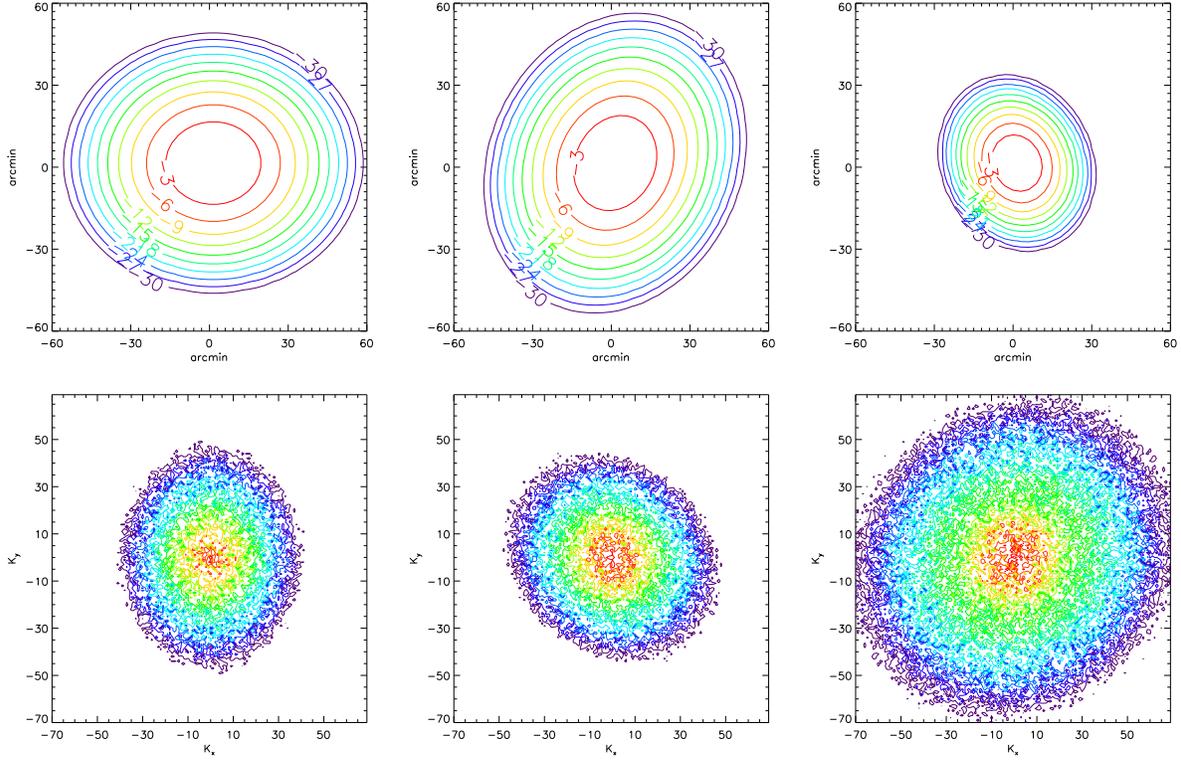,width=16cm}
\caption{Reciprocity of the beam in Real and Fourier space. Top row shows that the elliptical beam profile with the major and minor FWHM 36 and 30 arcmin (left and middle panel), and 21 and 18 arcmin (right) respectively, is used to convolve a $24\times24 {\rm deg}^2$ simulated CMB patch (without adding noise). The convolution is performed with a fixed orientation of $0^\circ$ (left), $60^\circ$ (middle), and $120^\circ$ (right) of the major axis against the x-axis. The bottom row shows the Fourier amplitude contour of the corresponding beam-convolved patches. We only show the amplitude contour down to $-30$ dB in order to display the shape and orientation. Note that the reciprocal property between the Real and Fourier space has caused the orientation manifested in the Fourier space a shift of $\pi/2$, and also the larger FWHM of the beam, the smaller it manifests itself in Fourier domain.}
\label{demo}
\end{figure}

In \citet{chiangbeam} it is demonstrated that the beam profile manifests itself in the 2D Fourier space. In Figure \ref{demo} we show 3 different orientations of the beam convolving on simulated CMB signal in a $24\times24 {\rm deg}^2$ patch and in the bottom row the corresponding Fourier amplitude of the beam-convolved patch. One of the important characteristics shown in the Fourier space is that the beam orientation turns $\pi/2$ due to the reciprocity between the Real and Fourier space. It is thus possible to extract the inflight effective beam information contained in the patches.

Below we present some methods that can reveal the inflight effective beam convolving the patch of the sky. 

\subsection{Direct effective beam estimate via bright extragalactic point sources}
\begin{figure}
\plotone{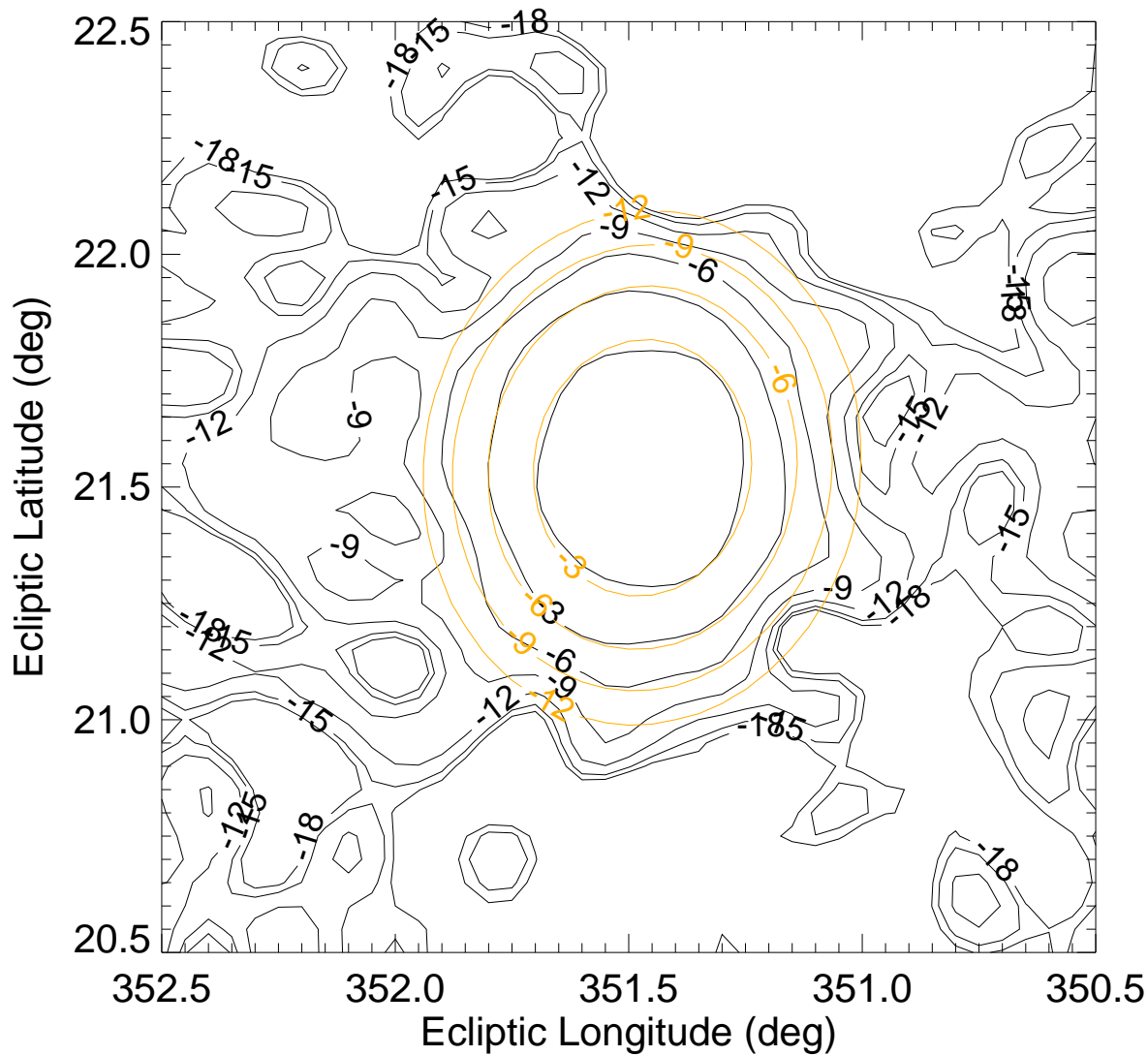}
\caption{For patches with bright point sources, we can estimate the asymmetry of the effective beam directly. The point source here is GB6 J2253$+$1608 appeared in Q1 DA. Direct estimate of its asymmetry and orientation results in $R=0.179$ and the orientation angle is $98^\circ.98$ for the major axis against Ecliptic Equator. We can also apply the Fourier method on the patches with point sources for consistency, which renders $R=0.110$ and angle $96^\circ.14$.}
\label{pointsource}
\end{figure}

For patches with bright radio point sources,  we can, following the assumption of slow rotation of the beam, further assume that the point source is the representative of the effective beam for the whole patch. Bright point sources are manifestation of the beam similar to the standard measurement of Jupiter for the beam profile, but that in point sources, we can get to at best $-10$ dB for most cases. Nevertheless, it is still useful for providing the asymmetry and orientation of the inflight effective beam. We show in Fig.\ref{pointsource} the bright point source GB6 J2253$+$1608 in Q1 DA as an example. The estimated asymmetry $R = 0.179$ and the angle is $98^\circ.98$ for the major axis against the (positive) Ecliptic Equator.

\subsection{Fourier method for estimation of the effective beam}
As shown in Fig.\ref{demo} the effective beam manifests itself in the Fourier domain, so we can extract the beam asymmetry and orientation directly from the Fourier amplitude. Assuming the foreground power spectrum $\propto k^{-2}$ \citep{wmap1yrfg}, we take the grid points ${\bf k}\equiv (k_x,k_y)$ in the Fourier domain where $|{\bf k}|^{-2}\exp(-|{\bf k}|^2\sigma^2) > r$,  $\sigma \equiv {\rm FWHM}/2\sqrt{2 \ln 2}$, and ${\rm FWHM}$ is the size of the nominal beam of the frequency band. Even for the same $r$, the number of grid points varies for different frequency bands due to different beam size. Also note that the level of $r$ cannot be lower than the pixel noise, thus simulation is required in order to get the optimal $r$. 

The other method is the phase perturbation method, developed in \citet{chiangbeam}. For the measured signal $S$ in Eq.(\ref{gen}), we can add the controlled white noise $W$ : $M = S + W $ to perturb the phases of the patch and calculate the phase shift $\Psi^M_{\bf k}-\Psi^S_{\bf k}$, where $\Psi^M_{\bf k}$ and $\Psi^S_{\bf k}$ are the Fourier phase at mode ${\bf k}$ for patch $M$ and $S$ respectively. Note that the controlled noise level cannot be lower than that of the pixel noise (the $N$ in Eq.(\ref{gen})). We can then calculate the mean from an ensemble of, say $n=200$, such perturbation:   
\begin{equation}
\Delta^2({\bf k})=\langle(\Psi^M_{\bf k} -\Psi^S_{\bf k})^2 \rangle|_n.
\end{equation}
For the Fourier modes where the Fourier amplitudes of the convolved signal are much higher than that of the controlled noise, the phases are not perturbed much, so the $\Delta^2({\bf k})\simeq 0$. When the controlled noise level is close to that of the signal, the phases are perturbed so much that can be approximated  
\begin{equation}
\Delta^2({\bf k})\simeq\frac{\langle |W_{\bf k}|^2\rangle}{2|S_{\bf k}|^2}, 
\end{equation}
where the controlled noise level is set as the same as that of the pixel noise \citep{chiangbeam}.

\section{Demonstration and accuracy and consistency of the Fourier method} 
\begin{figure}
\centering
\epsfig{file=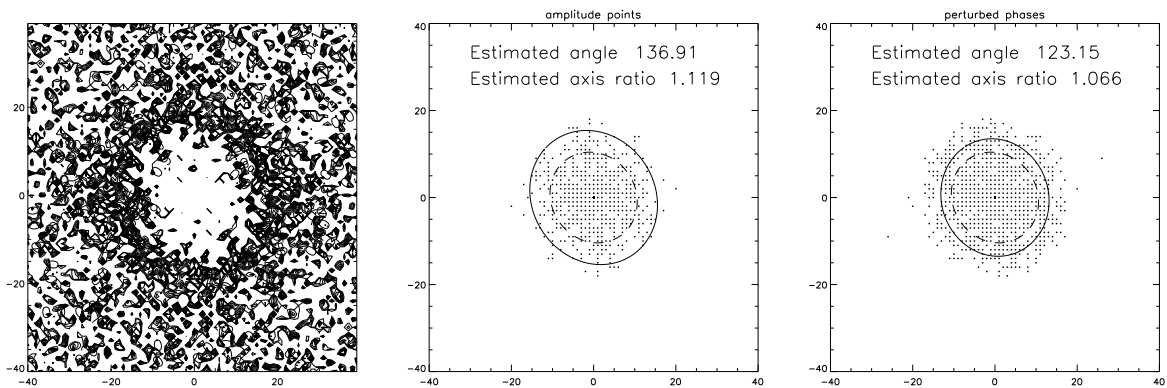,width=16cm}
\caption{Demonstration of the estimate of the effective beam from the Fourier method. The left panel shows the contour of the Fourier power (mK$^2$) up to $10^{-6}$, the level equivalent to around $-5$ dB of any beam on the best-fit from the input beam $R=0.1$ and major axis angle $50^\circ$ convolving the simulated CMB signal. Note that one can see clearly the asymmetry and the major axis in the Fourier space is shift by $90^\circ$ due to the reciprocal property between real and Fourier space. Middle and Right: estimation from the points of the Fourier power collected down to $-5$ dB and from phase perturbation method, respectively. The dash ellipse is the input shape and solid one the estimated. We make the input and output ellipse different sizes in order to show how tiny the difference is.}
\label{simu}
\end{figure}

To demonstrate the Fourier method for estimation of the effective beam, an elliptical beam whose FWHM of major and minor axis is 33 and 30 arcmin, respectively (thus asymmetry $R = 0.1$), is used to convolve a $24\times24 \;{\rm deg}^2$ CMB map simulated with the best-fit $\Lambda$CDM model before adding noise at level $\sigma=0.152$ mK. The orientation of the major axis is $50^\circ$ against the $x$-axis. We also make the patch with non-periodic boundary to mimic the real situation. We then take the grid points $> -5$dB for amplitude method and those with $\Delta< 9^\circ$ for phase perturbation method, and fit them with free-parametered elliptical shape. As shown in Fig.\ref{simu}, once again the reciprocal property causes the orientation of the beam turned $\pi/2$ in the Fourier space, now $140^\circ$. Note that in our simulation we use a Gaussian elliptical function to mimic the effective beam, and omit the collective effect from the far sidelobe of the beam. We then employ the method described above and the estimated asymmetry and orientation from the amplitude information is $R =0.119$ and $136^\circ.9$, and $R = 0.066$ and $123^\circ.2$ from phase perturbation method, which is to be compared with the input $R =0.1$ and orientation angle $140^\circ$. One can see that the error from the angle is $-3^\circ.01$ and $-16^\circ.8$, respectively, whereas the error of the asymmetry is 0.019 and $-0.034$.

One source of error in such estimation comes from small number of points on equal-spaced grid in Fourier domain. The less the points available in Fourier space, the higher error in estimation. The number of points available from the Fourier grid is limited by the beam size and the noise level. For estimating small beam size such as \wmap W band, there are more points in the Fourier grids (due to reciprocity), but the noise level is higher than the low-frequency bands. For larger beam in K band, there are less points. Thus to gain more available grid points we resort to large patches such that the white noise level can be lowered.

\begin{figure}
\epsfig{file=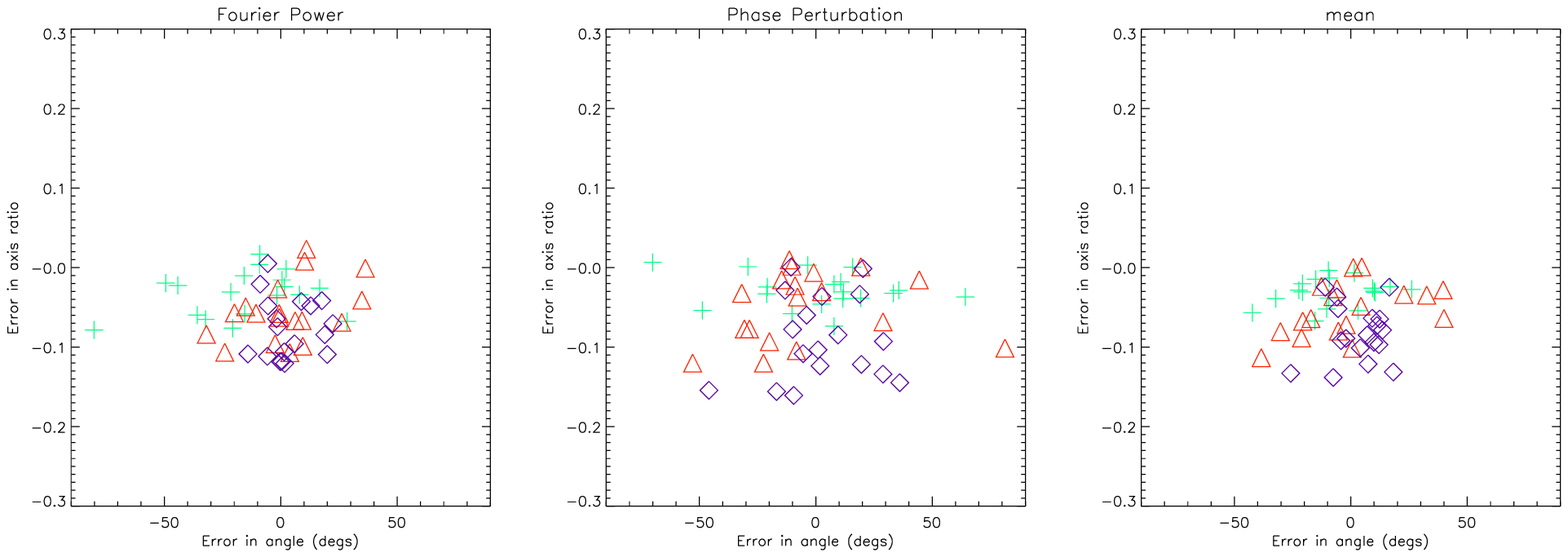,width=16cm}
\caption{We test the accuracy and consistency of the two Fourier methods on various beam parameters. We show the error from the estimation compared to the simulation input. In all 3 panels the plus, triangle and diamond sign denotes $r_0=1.10$, 1.15, and 1.20, respectively. the Left panel shows the estimate from Fourier amplitude method : error in asymmetry $\langle \Delta r \rangle/r_0 =−0.055\pm 0.038$, error in angle $\langle \Delta\theta \rangle= −3.15\pm 21.08$ deg. Middle is that from phase perturbation method:  $\langle\Delta r\rangle /r_0=−0.056\pm 0.049$, $\langle \Delta\theta\rangle= −0.05\pm 27.74$ deg. And for the mean from both estimate, the error is shown in the right panel $\langle \Delta r\rangle /r_0=−0.056\pm 0.036$, $\langle \Delta\theta \rangle= −1.60\pm 18.05$ deg.}
\label{validity}
\end{figure}

We conduct simulations on \wmap Q band with the nominal beam size FWHM 30.6 arcmin for 3 different ratios $r_0 =1.10$, 1.15 and 1.20 \footnote{Here we use the defined $r$, the ratio of the axes, instead of asymmetry $R$ because the error ratio shall be defined in terms of the $r$, not $R$.}, each with 18 different orientations. We apply the two Fourier methods to test their accuracy and consistency. In Fig.\ref{validity} we plot the error in asymmetry against the error in orientation angle for the Fourier amplitude method (left panel) and phase perturbation method (middle) on the simulations. The Fourier amplitude method skews the estimate in orientation angle for $r_0 =1.10$ (plus sign), but has better accuracy and consistency for $r_0 =1.15$ (triangle sign) and 1.20 (diamond sign). Phase perturbation method has more accuracy but less consistency. Since each use different information from the Fourier space (though not independently), we take the mean from both estimates and plot on the right panel. The estimated error in 
angle is $\langle \Delta\theta \rangle = −1.60 \pm 18.05$ deg and error in axis ratio $\langle \Delta r \rangle/r_0 = −0.056\pm0.036$, indicating our estimate on the orientation angle is reasonably good, while the asymmetry is underestimated by 5.6\%. 

We can also use the Fourier method mentioned above for the patch containing the bright point source GB6 J2253$+$1608 (Fig.\ref{pointsource}). The asymmetry from Fourier method is $R = 0.110$ and angle is $96^\circ.14$. The difference in the estimated asymmetry $-0.069$ is as predicted: our Fourier method consistently underestimate the asymmetry, whereas the difference in the estimated angle $-2^\circ.84$ is negligible.   
 
\section{Asymmetry and non-random orientation of the inflight effective beam of the \wmap data} 
We can now employ the methods we demonstrate in the previous section to estimate the inflight effective beam on the \wmap data. We divide the whole map into $24^\circ \times 24^\circ$ patches and apply the Fourier method described above on the patches. The reason why we estimate the beam on such a large patch is that the noise level in the \wmap data is quite high, which inevitably set the limit of the available Fourier grid points. We show our estimate of the inflight effective beam of the \wmap 9-year Q1 DA. 

\begin{figure}
\epsfig{file=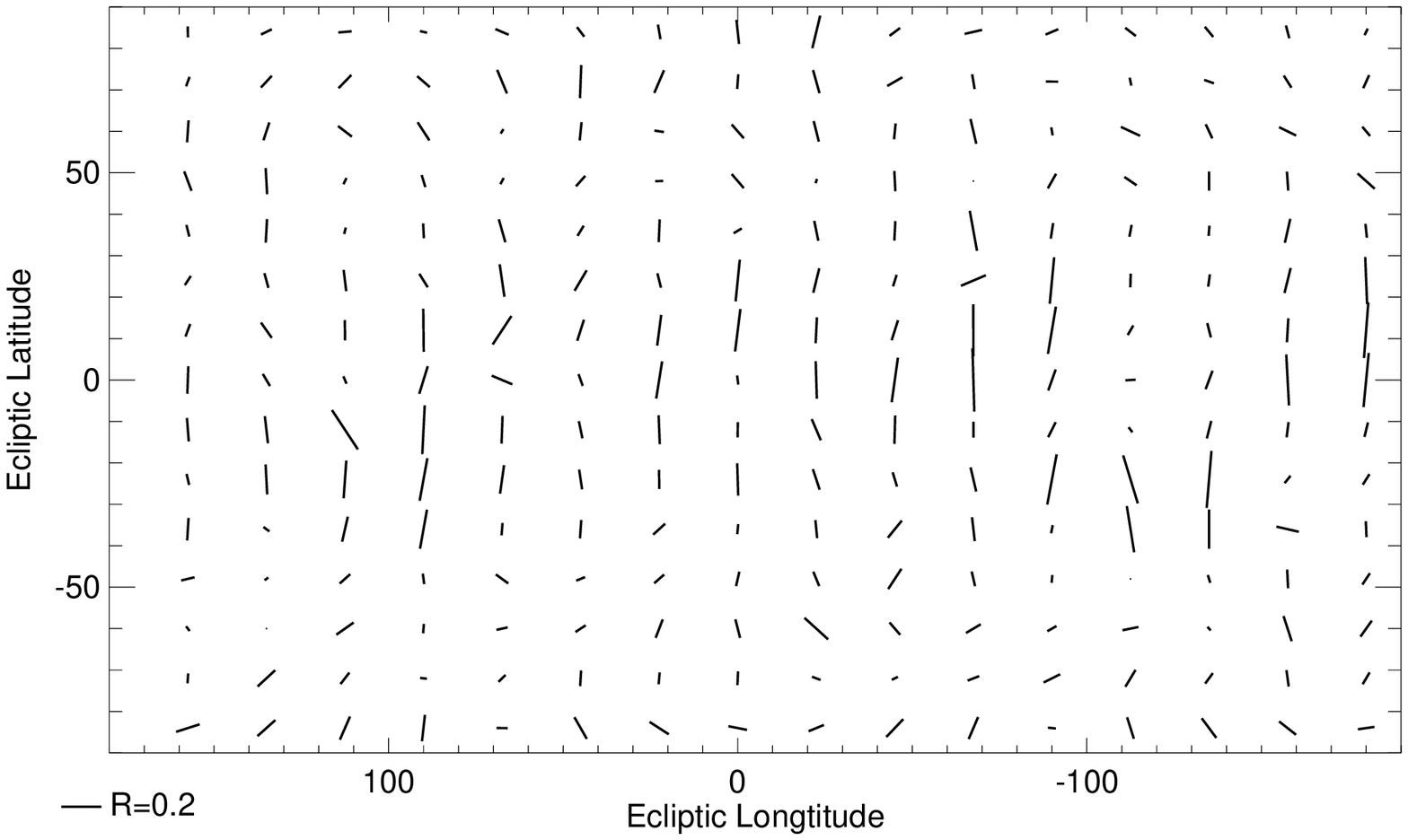,width=15cm}
\epsfig{file=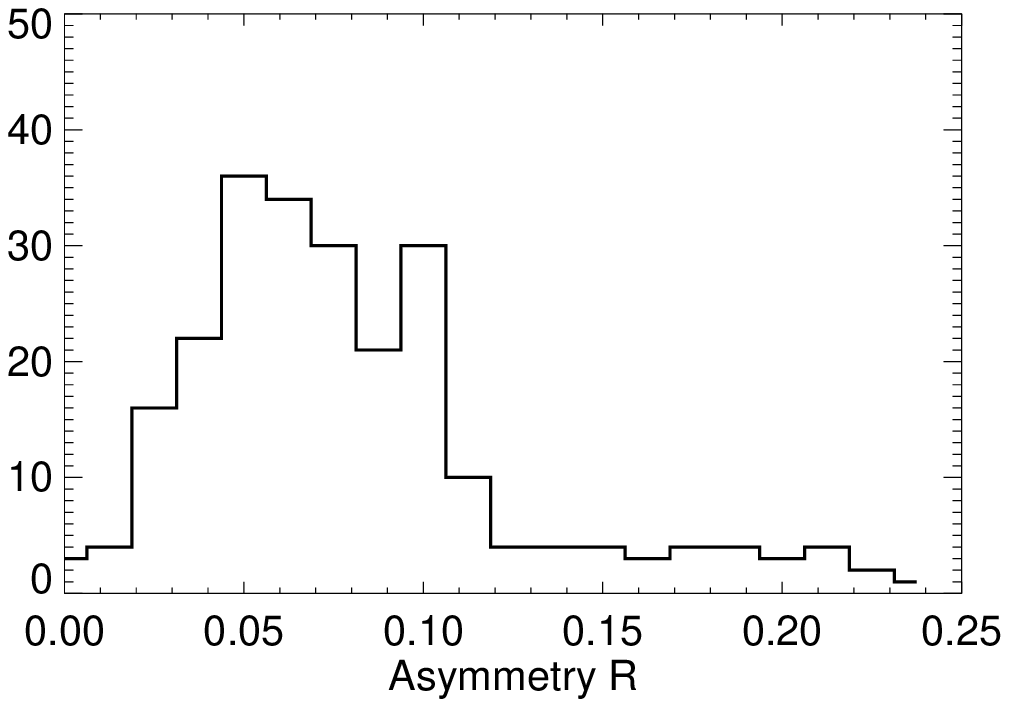,width=7cm}
\epsfig{file=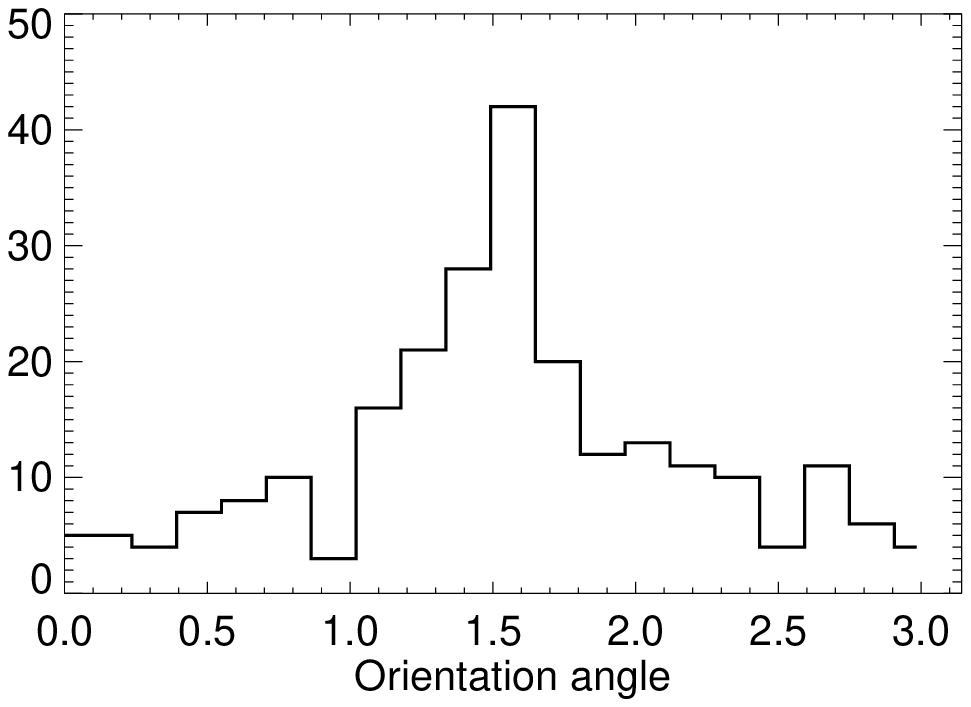,width=7cm}
\caption{Estimation of the effective beam of \wmap Q1 DA in Ecliptic Coordinate. The bar denotes the asymmetry R and the orientation of the major axis is denoted by that of the bar. The bottom two panels show the histogram of the asymmetry $R$ and the orientation angle. The angle is defined with the that between the bar and the Ecliptic Equator.}
\label{q1}
\end{figure}

\begin{figure}
\epsfig{file=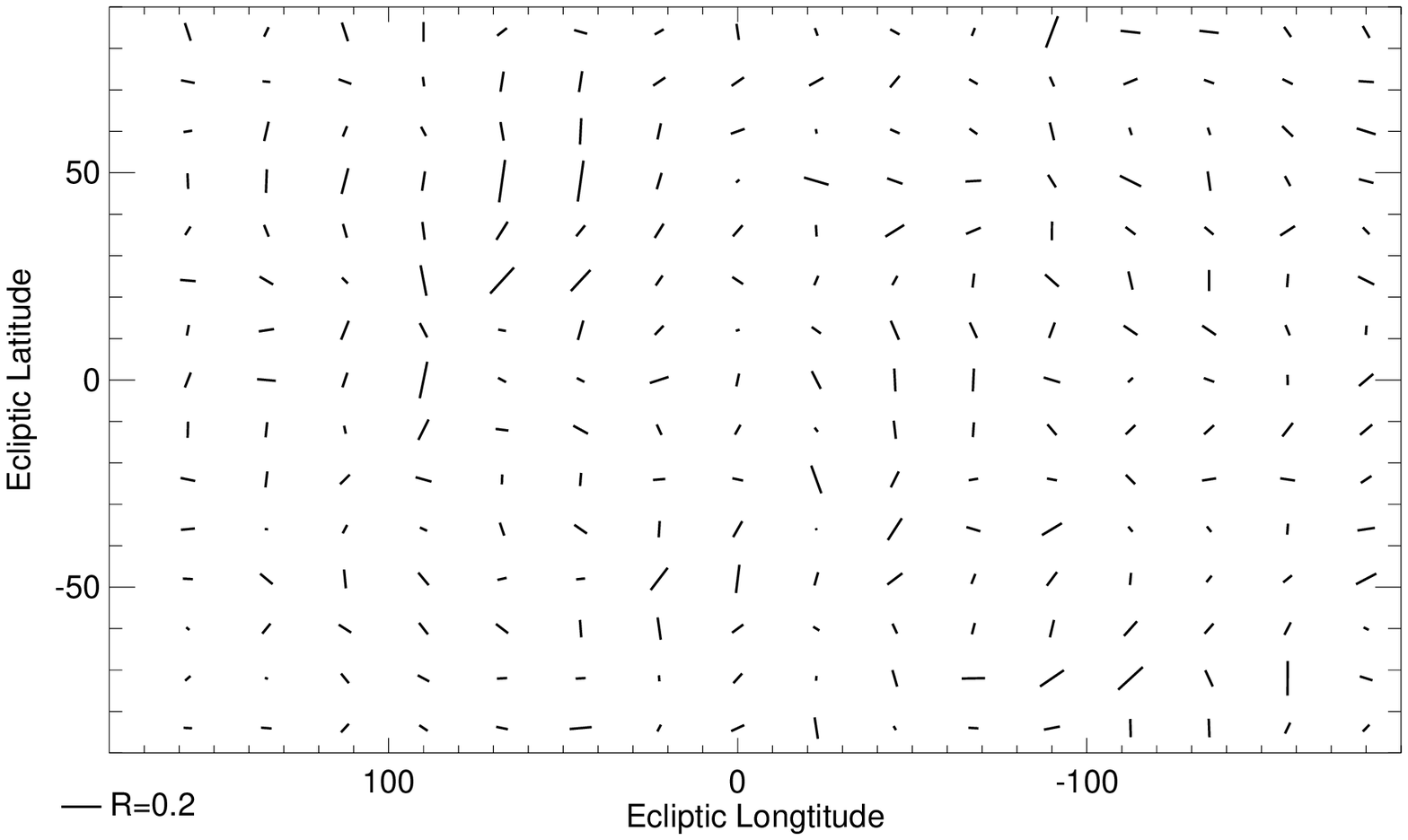,width=15cm}
\epsfig{file=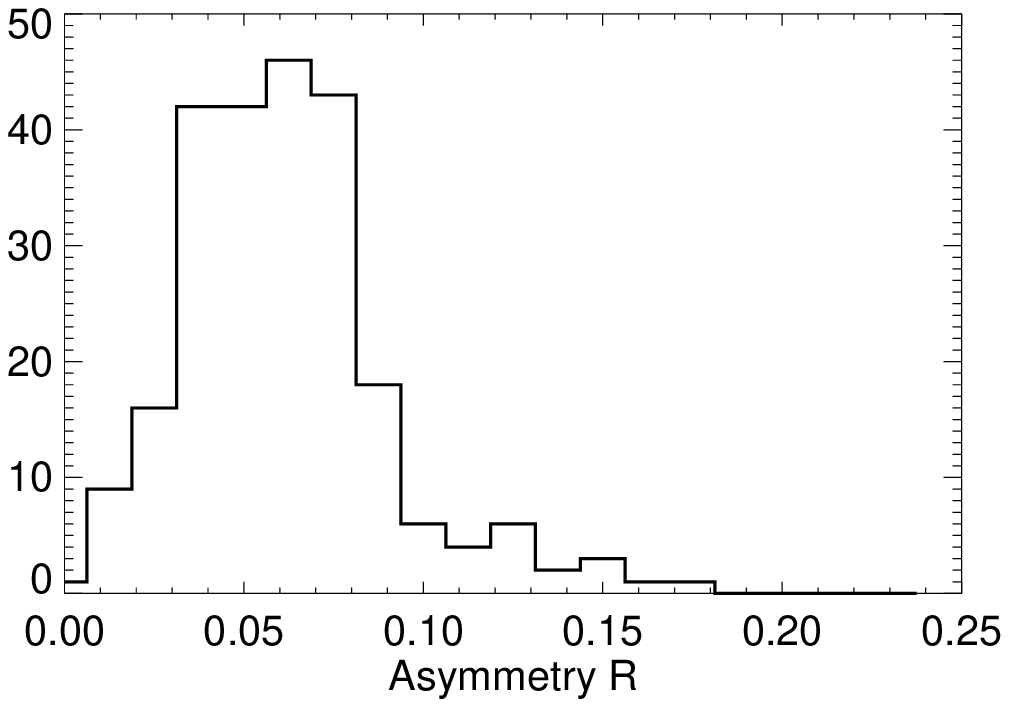,width=7cm}
\epsfig{file=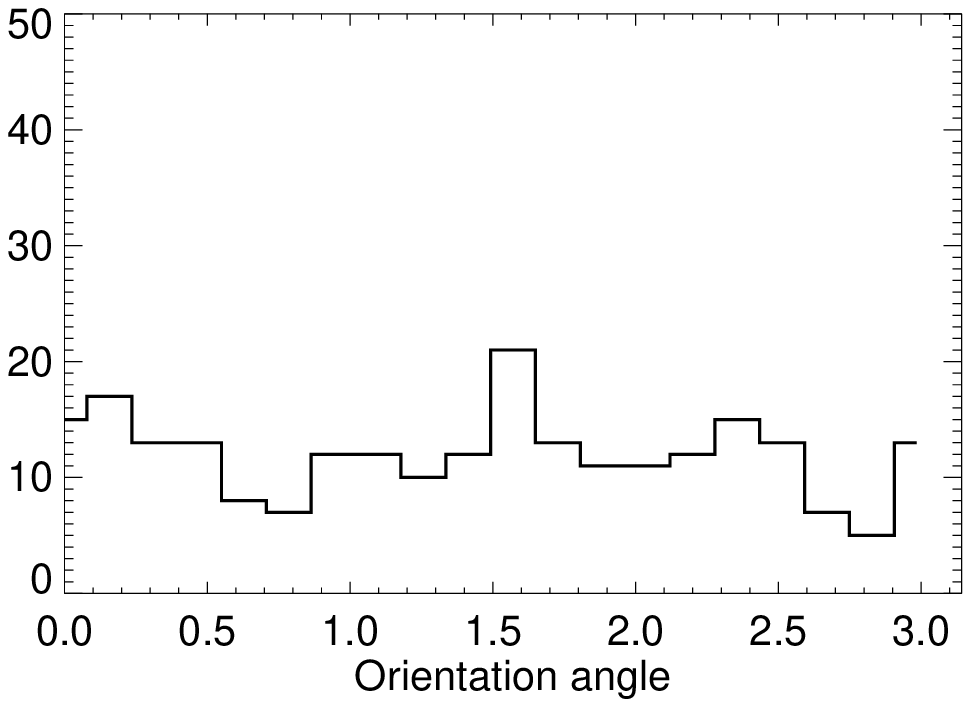,width=7cm}
\caption{Estimation of the effective beam of \wmap ILC map in Ecliptic coordinate. All the notations are the same as in Fig.\ref{q1}.}
\label{ilc}
\end{figure}

Due to the scan strategy that the \wmap observes from a Lissajous orbit about the L2 Sun-Earth Lagrange point, and telescope line of sight is around $70^\circ$ off the \wmap spinning axis, the path swept out on the sky by a given line of sight resembles a Spirograph pattern that reaches from the north to south ecliptic poles, hence the inflight beam pattern is closely related to the Ecliptic coordinate. In Fig.\ref{q1} we plot the estimated inflight effective beam pattern. The length of the bars indicates the asymmetry in $R$ and the inclination denotes that of the major axis. The bottom 2 panels show the histogram of the asymmetry $R$ and the orientation angle. The angle is defined with the that between the bar and the Ecliptic Equator. We plot the same for the \wmap ILC map in Fig.\ref{ilc}. 

Our results confirm that the inflight effective beams convolving the underlying signal are asymmetric for most parts of the sky, and are not randomly oriented. The alignment is most severe around Ecliptic Equator. Near the ecliptic poles, however, the asymmetry is small due to the averaging effect from different orientations from the scan strategy. The histogram of the orientation angle shows high concentration around $\pi/2$. On the other hand, the ILC map results from the combination of different frequency band maps, which have different beam orientations and beam sizes, hence the asymmetry the orientation alignment of the effective beam on the ILC map is ameliorated by the internal combination.

\section{Conclusion and Discussion} 
In this paper we have developed a method to reveal the asymmetry and orientation of the inflight effective beam in the \wmap data. We divide the whole sky map into patches and exploit the information resided in the Fourier domain. We test the accuracy and consistency of our Fourier method with simulations and also with the bright point source in the patch, the shape of the latter being a representative of the inflight effective beam. We then apply the method on the \wmap Q1 DA and ILC map, and it is confirmed that the effective beam is rather asymmetric with strong alignment in their orientation around the Ecliptic Equator in the Q1 DA map. That the asymmetry of the effective beam in the ILC map is lessened does not guarantee the ILC map is not without the systematic error from the beam effect. With highly-aligned beam with significant asymmetry combing through the measured signal, which includes strong emission of the galactic foreground, can have elongated effect on the signal, which shall cause some systematic error in the foreground cleaning process for the CMB if it is not carefully treated. The method developed in this paper can be used on the Planck data.

\acknowledgments
We acknowledge the use of \healpix\footnote{{\tt http://healpix.jpl.nasa.gov/}} package \citep{healpix}  and the use of \glesp\footnote{{\tt http://www.glesp.nbi.dk/}} package \citep{glesp}. The author would like to thank Pavel Naselsky for useful discussions and suggestions.

\expandafter\ifx\csname natexlab\endcsname\relax\def\natexlab#1{#1}\fi
\newcommand{\combib}[3]{\bibitem[{#1}({#2})]{#3}} 

%
%
\newcommand{\autetal}[2]{{#1,\ #2., et al.,}}
\newcommand{\aut}[2]{{#1,\ #2.,}}
\newcommand{\saut}[2]{{#1,\ #2.,}}
\newcommand{\laut}[2]{{#1,\ #2.,}}
\newcommand{\coll}[1]{#1,}

%
%
\newcommand{\refs}[6]{#5, #2, #3, #4} 
\newcommand{\unrefs}[6]{#5 #2 #3 #4 (#6)}  

%
%

\newcommand{\book}[6]{#5, #1, #2, #3}
%

\def\apj{ApJ}
\def\apjl{ApJL}
\def\mn{MNRAS}
\def\nature{Nature}
\def\aa{A\&A}
\def\aas{A\&A Supplement}
\def\prl{Phys.\ Rev.\ Lett.}
\def\prd{Phys.\ Rev.\ D}
\def\pr{Phys.\ Rep.}
\def\ijmpd{Int. J. Mod. Phys. D}
\def\jcap{J. Cosmo. Astropar.}

\end{document}